\documentclass[preprint,12pt]{elsarticle}

\usepackage{amssymb}
\usepackage{amsmath}

\usepackage{lineno}

\journal{Nuclear Instruments and Methods in Physics Research Section A}

\usepackage{amsthm,mathrsfs,algorithmicx}
\usepackage{textcomp,multirow,booktabs,xcolor,tabularx,colortbl}
\usepackage{graphicx,pgfplotstable,pgfplots,subcaption}
\usepackage{isotope}
\usepackage[separate-uncertainty=true, per-mode=single-symbol]{siunitx}
\usepackage[title]{appendix}
\usepackage[T1]{fontenc}
\usepackage{hyperref}

\pgfplotsset{compat=1.18}

\newcommand{\eq}{\; = \;}

\begin{document}

\begin{frontmatter}



\title{Detecting Muons with a Siemens Biograph Vision Quadra} 

\author[aff1,aff2]{Lorenzo Mercolli\corref{cor1}}
\author[aff3]{William M. Steinberger}

\cortext[cor1]{Corresponding author: lorenzo.mercolli@unibe.ch}

\affiliation[aff1]{organization={Department of Nuclear Medicine, Inselspital, Bern University Hospital, University of Bern},
            addressline={Rosenb{\"u}hlgasse 25},
            city={Bern},
            postcode={CH-3010},
            country={Switzerland}}
\affiliation[aff2]{organization={Albert Einstein Center for Fundamental Physics (AEC), Laboratory for High Energy Physics (LHEP), University of Bern},
            addressline={Sidlerstrasse 5},
            city={Bern},
            postcode={CH-3012},
            country={Switzerland}}
\affiliation[aff3]{organization={Siemens Medical Solutions USA, Inc.},
            addressline={810 Innovation Dr.},
            city={Knoxville},
            postcode={37932},
            state={TN},
            country={USA}}

\begin{abstract}
Atmospheric muons are routinely detected with a variety of instruments, from cloud chambers to scintillation detectors. In this study, we demonstrate that atmospheric muons can also be detected with a clinical long axial field-of-view positron emission tomograph (PET). We performed a blank scan without any radioactive source, using a prototype feature that records all single crystal interactions. The energy a muon deposits in the PET crystals far exceeds the upper limit of the recorded energy range, which means muons cannot be identified through energy discrimination. Instead, we used the scanner’s time-of-flight capabilities to select crossing events, i.e.\ pairs of high-energy single crystal events across the PET gantry whose spatial and temporal separation is consistent with a downward-going ultrarelativistic particle. Random pairings, however, cannot be rejected on an event-by-event basis. We therefore established the muon origin of the crossing events through their characteristic angular distribution. In order to account for geometry and event selection bias, we compare the measured crossing events to a Monte Carlo simulation of muons following the expected angular distribution. This confirms that the measured crossing events in the PET scanner are consistent with atmospheric muons. 
\end{abstract}

\begin{highlights}
\item First detection of atmospheric muons with a commercial PET/CT scanner.
\item TOF-based pairing of high-energy gantry-crossing events.
\item Angular distribution of gantry-crossing events agrees with expected distribution. 
\end{highlights}


\end{frontmatter}

\section{Introduction}

\begin{figure}
    \centering
    \includegraphics[width=0.5\linewidth]{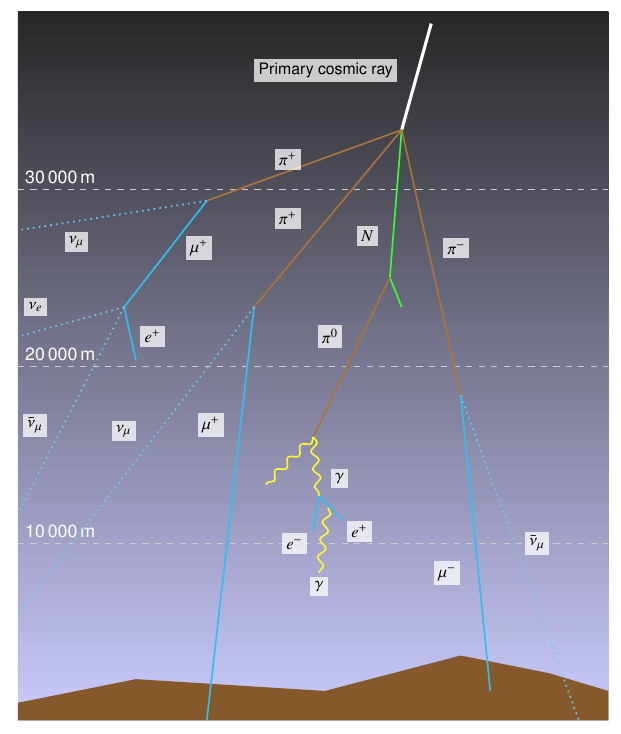}
    \caption{Schematic illustration of $\mu$ production in the upper atmosphere. }
    \label{fig:atmospheric_muons}
\end{figure}

Cosmic rays — mostly protons, with a smaller admixture of heavier nuclei — strike the upper atmosphere and initiate a cascade of secondary particles through interactions with nuclei (see e.g.\ review on cosmic rays in Ref.~\cite{PDG2024}). Muons, i.e.\ charged leptons with a mass of \SI{105.6583755\pm0.0000023}{\MeV}, observed at the Earth's surface originate from the decay of the pions and kaons generated in these primary interactions as illustrated schematically in Fig.~\ref{fig:atmospheric_muons}. At sea level, the integral intensity of vertical muons is about $I \approx \SI{1}{\per\square\cm\per\minute}$, with a mean kinetic energy around \SI{4}{\giga\electronvolt} \cite{PDG2024}. Given the large kinetic energy of these charged particles, muons have been used to study large natural or man-made structures, with application across various disciplines such as particle physics, geology, archaeology, homeland security or nuclear waste \cite{Bonechi2020}. The detection of atmospheric muons is also a standard classroom experiment in most physics curricula \cite{DesyCosmoMill}. 

Muons can be detected with a wide variety of detectors for ionizing radiation, such as cloud chambers, cathode strip chambers, scintillation based detectors, etc. In the present study we investigate the possibility of detecting atmospheric muons with a commercial positron emission tomography (PET) scanner. PET in combination with computed tomography (CT) has become an indispensable medical imaging modality with application in oncology, neurology and cardiology \cite{Farwell2014,Conti2019,Duclos2021,Meikle2021}. PET/CT scanners are optimized for detecting coincident photon pairs that stem from an electron-positron annihilation process inside a patient's body. Detecting muons with such a device requires going beyond the standard capabilities of a clinical PET/CT scanner \cite{Zhao2026}. 

In this study we show how to detect atmospheric muons with a Biograph Vision Quadra (Siemens Healthineers, Knoxville, TN, USA) \cite{Prenosil2022}. Quadra is a long axial field-of-view (LAFOV) PET/CT that is used in clinical routine throughout the world. In recent years, Quadra has been used as a platform to push the envelope of what commercial PET/CT scanners can do, as e.g.\ in the case of positronium lifetime imaging \cite{Steinberger2024,Mercolli2025a,Mercolli2026,Mercolli2026a} or exploiting the natural abundance of \isotope[176]{Lu} in the crystals for attenuation maps \cite{Teimoorisichani2022,Sari2022} and automatic quality control \cite{Rothfuss2014}. In the same context, Ref.~\cite{Porcelli2025} recently proposed to investigate atmospheric muons with a prototype PET scanner.

While we do not envision a direct medical application of detecting muons with a commercial PET/CT, we could see the methodology of our study applied in different areas. E.g.\ the muon signal could be used for automatic quality control without radioactive sources, in particular for PET scanners that do not have radiation from LSO crystals. Also, considering the spread of PET/CT scanners across the globe, it is conceivable that muon fluxes could be detected synchronously at different locations. (see also the proposal in Ref.~\cite{Porcelli2025}).

\section{Methods}

\subsection{Instrumentation}

The basic detector elements of Quadra are the same as for the Biograph Vision 600 (Siemens Healthineers, Knoxville, TN, USA) \cite{Sluis2019,Prenosil2022}. A mini-block (MB) is a $5\times5$ array of $3.2\times 3.2\times\SI{20}{\mm\cubed}$ large lutetium oxyorthosilicate (LSO) crystals. Two rows of $4\times16$ MB, with a small angular displacement, constitute a detector electronics assembly (DEA). Each of the four detector rings of Quadra is made of 19 DEAs \cite{Prenosil2022}. Importantly, Quadra has time-of-flight (TOF) capabilities with a TOF resolution of \SI{228}{\pico\s} at \SI{5.3}{\kilo\becquerel\per\ml} with a maximum full-ring difference of 322 crystal rings \cite{Prenosil2022}. Ref.~\cite{Prenosil2022} determined also the energy resolution of Quadra to be \SI{9.8}{\percent} at \SI{511}{\kilo\electronvolt}. The axial field-of-view (FOV) measures \SI{106}{cm} and the gantry has a diameter of \SI{82}{cm}. 

Like every other PET/CT system, Quadra is optimized for the detection of annihilation photons around \SI{511}{\keV}. Typically, a low-lying trigger system sorts the coincidence photons and the list-mode data output of PET systems only contains information for coincident events. Quadra, however, features a prototype acquisition mode, called \emph{singles mode}, that allows users to write all single crystal interactions to a file. Refs.~\cite{Steinberger2024,Mercolli2025a,Mercolli2026a} showed the capability of singles mode for finding three-photon events to measure positronium lifetimes with a clinical PET/CT. 

Singles mode provides the location, timing and energy of a crystal interaction. Ref.~\cite{Steinberger2024} found a timing resolution of single crystal events to be slightly below \SI{0.1}{\nano\s}. As discussed in Refs.~\cite{Steinberger2024,Mercolli2025a,Mercolli2025}, the recorded energy of a single crystal event is limited to \SI{726}{\keV}. Quadra does detect crystal events with higher energies, but they all get assigned the same energy of \SI{726}{\keV}, i.e.\ this is an integrating energy bin (see also Ref.~\cite{Mercolli2025} for a discussion of negative effects of the integrating bin in the context of positronium lifetime imaging). 
In addition, singles mode still imposes scatter recovery between mini-blocks. Multiple hits in adjacent crystals are integrated over four MB. 

For the study of muons, we performed a \SI{20}{\min} blank scan, i.e.\ without any radioactive or positron source, in singles mode at the Department of Nuclear Medicine of Inselspital, Bern University Hospital in Switzerland. The resulting list mode file is about \SI{80}{\giga\byte} in a highly compressed binary format.

\subsection{Event selection}

\begin{figure}
    \centering
    \begin{subfigure}{0.17\textwidth}
        \includegraphics[width=\linewidth, trim={20cm 7cm 25cm, 6cm}, clip]{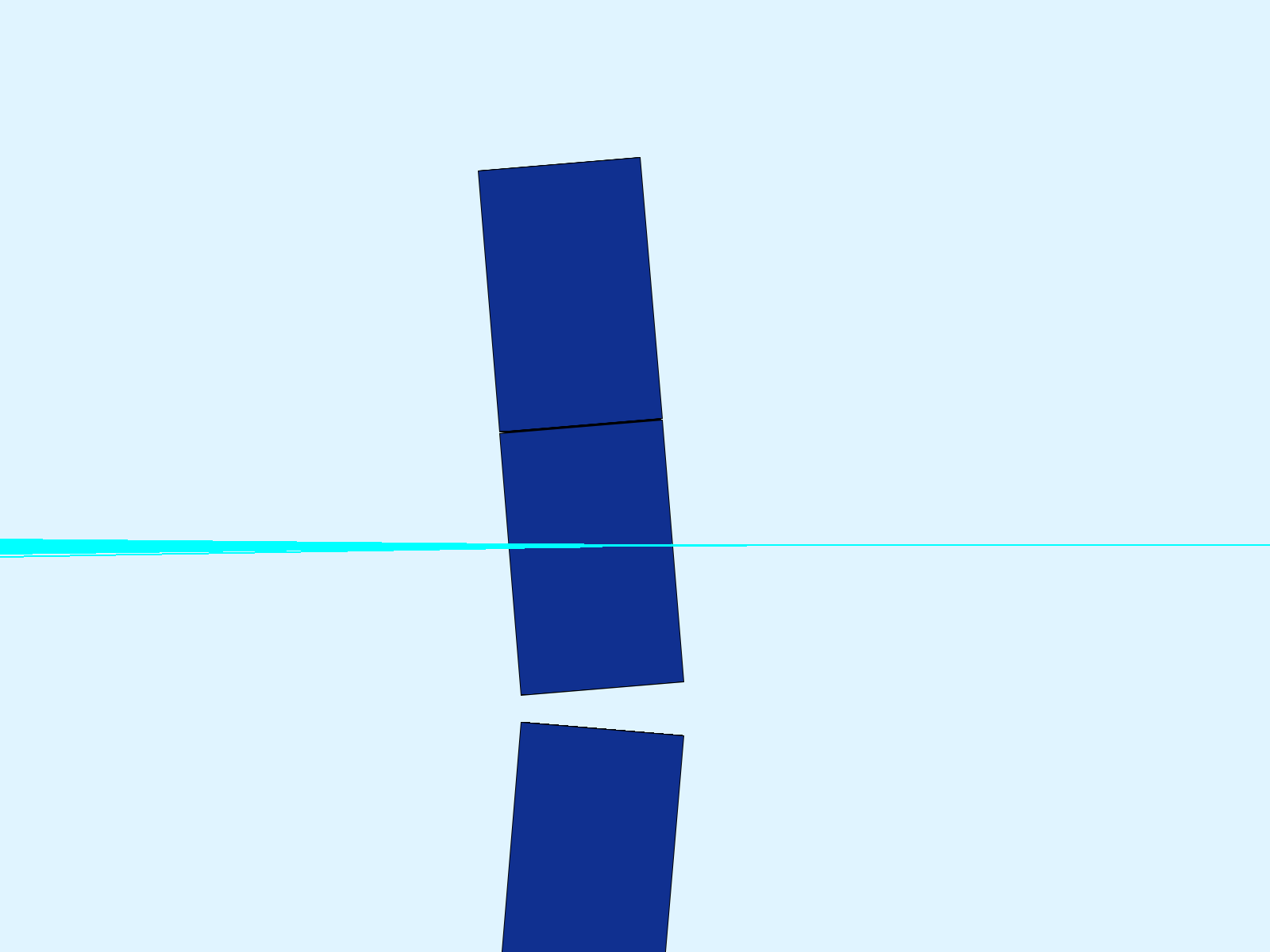}
        \caption{} \label{fig:deatracks_mu}
    \end{subfigure}
    \hspace{0.8cm}
    \begin{subfigure}{0.17\textwidth}
        \includegraphics[width=\linewidth, trim={20cm 7cm 25cm, 6cm}, clip]{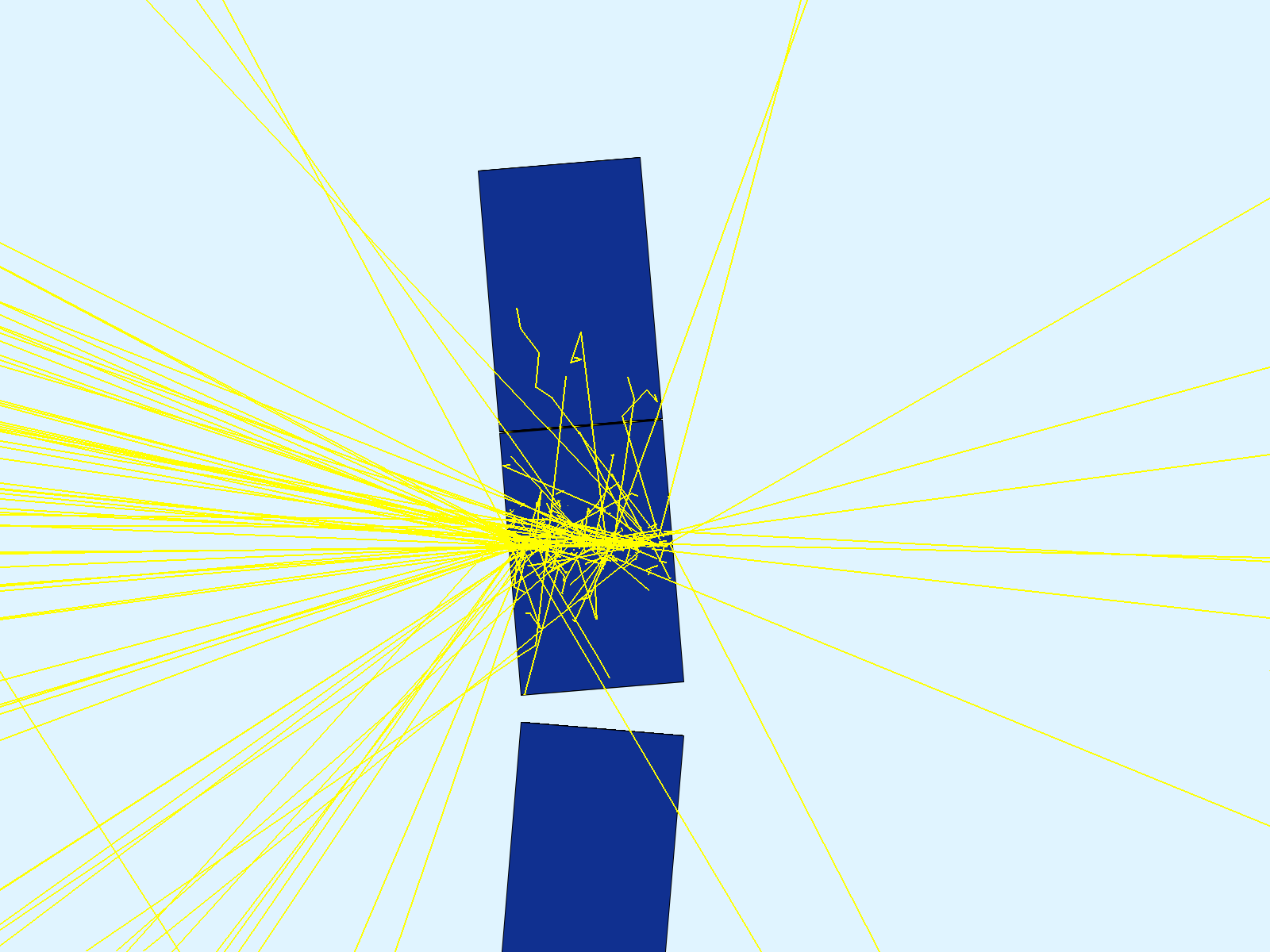}
        \caption{} \label{fig:deatracks_gamma}
    \end{subfigure}
    \hspace{0.8cm}
    \begin{subfigure}{0.17\textwidth}
        \includegraphics[width=\linewidth, trim={20cm 7cm 25cm, 6cm}, clip]{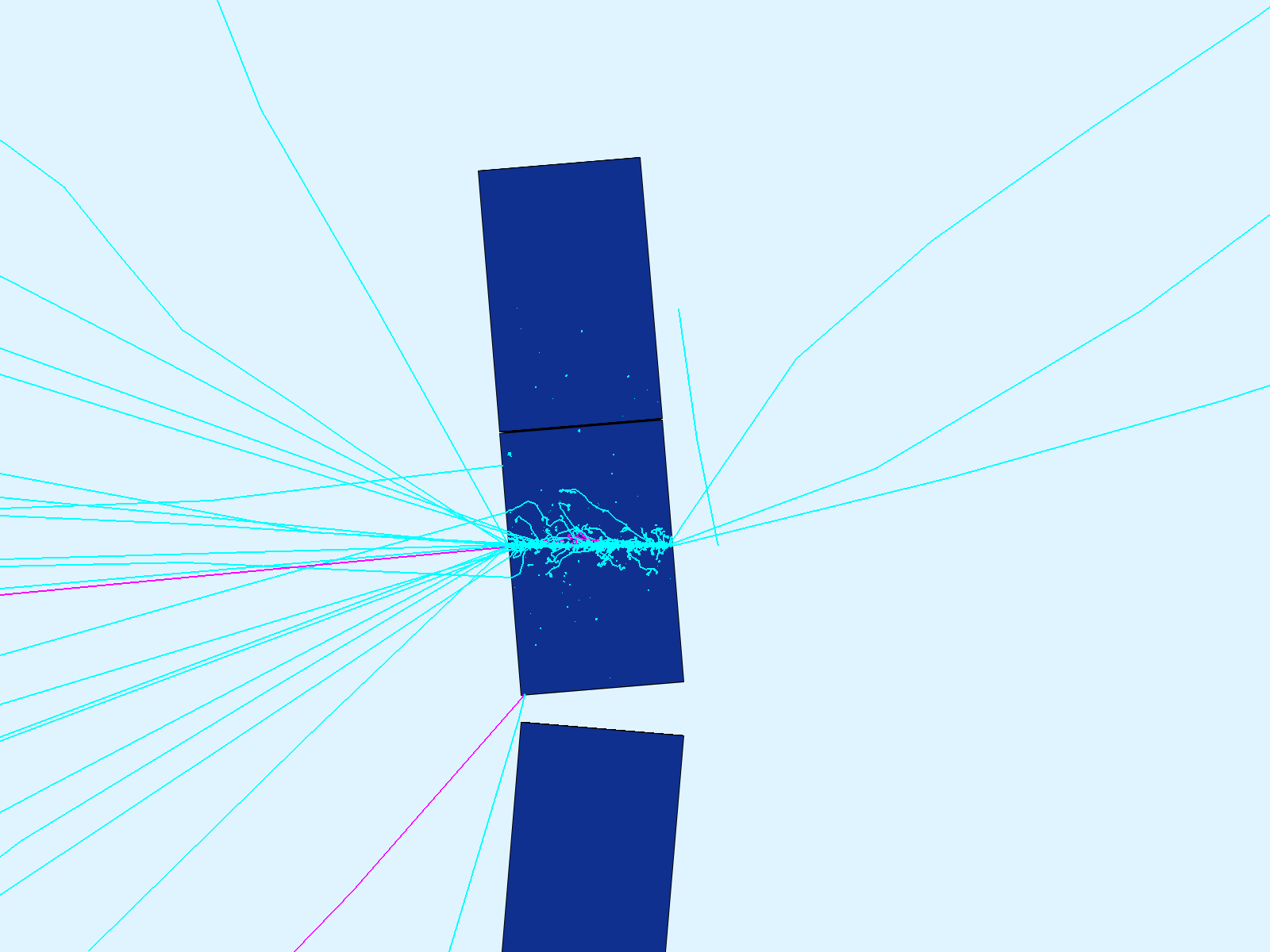}
        \caption{} \label{fig:deatracks_e}
    \end{subfigure}
    \caption{Simulated particle tracks of muons (\subref{fig:deatracks_mu}), secondary photons (\subref{fig:deatracks_gamma}) and electrons/positrons (\subref{fig:deatracks_e}, cyan/magenta ) crossing a DEA in the transaxial direction.} \label{fig:deatracks}
\end{figure}

\begin{figure}
\centering
    \begin{subfigure}{0.49\textwidth}
        \includegraphics[width=\textwidth]{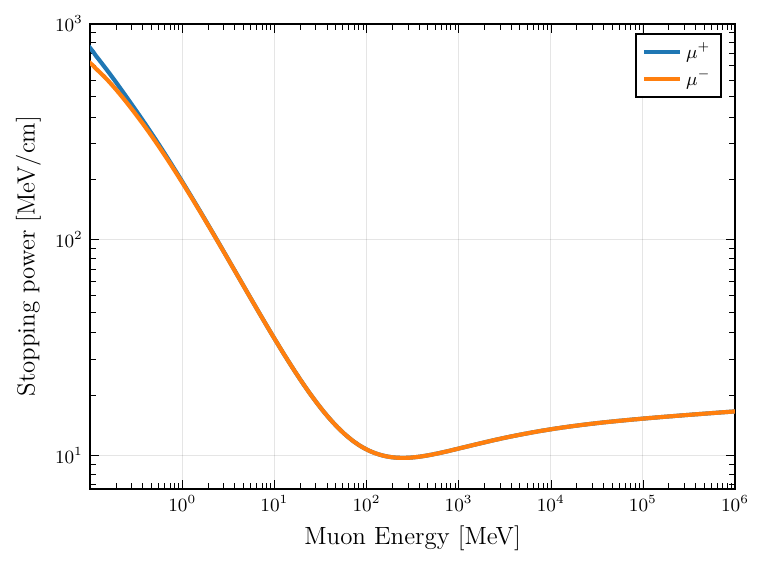}
        \caption{} \label{fig:stoppingpower}
    \end{subfigure}
    \hfill
    \begin{subfigure}{0.49\textwidth}
        \includegraphics[width=\textwidth]{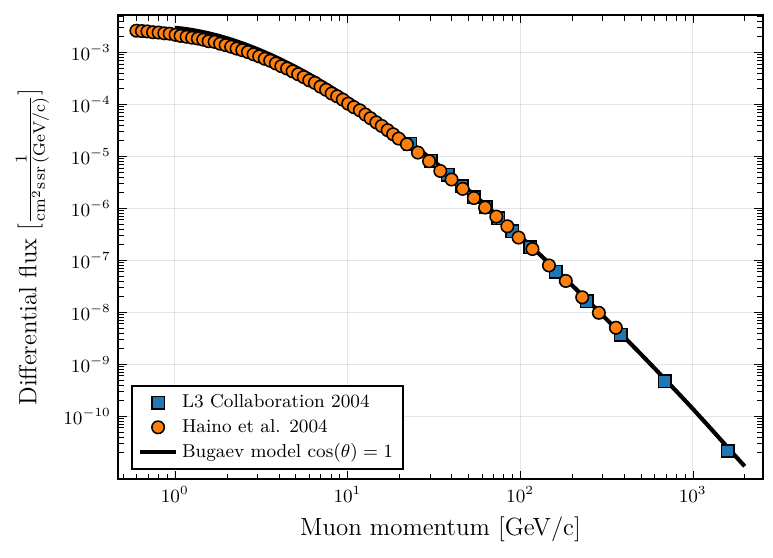}
        \caption{} \label{fig:muspectrum}
    \end{subfigure}
    \caption{Muon stopping power in LSO according to FLUKA (\subref{fig:stoppingpower}). Measurement data \cite{L32004,Haino2004} and Bugaev parametrization \cite{Bugaev1998} for the spectrum of vertical muons at sea level.}  \label{fig:stoppingpower_spectrum}
\end{figure}

From the singles mode output of Quadra, we have to select the events that could signal the passage of an atmospheric muon through the PET scanner. From the muon stopping power in LSO shown in Fig.~\ref{fig:stoppingpower} and the muon spectrum at sea level for vertical muons in Fig.~\ref{fig:muspectrum}, one can infer that the energy deposited by a muon in the LSO crystals is likely to be much larger than \SI{726}{\keV}. Fig.~\ref{fig:deatracks} illustrates the particle tracks of muons traversing a single DEA and secondary particles (photons, electrons and positrons) from a FLUKA simulation \cite{fluka1,fluka2,fluka3,flair}. 
This means that all muon events will end up in the high-energy bin and it is therefore not feasible to select muons impinging Quadra through energy discrimination. Also, we have no means to distinguish a single crystal event stemming from a muon from one stemming from another high-energy secondary particle. 

\begin{figure}
    \centering
    \includegraphics[width=0.55\linewidth]{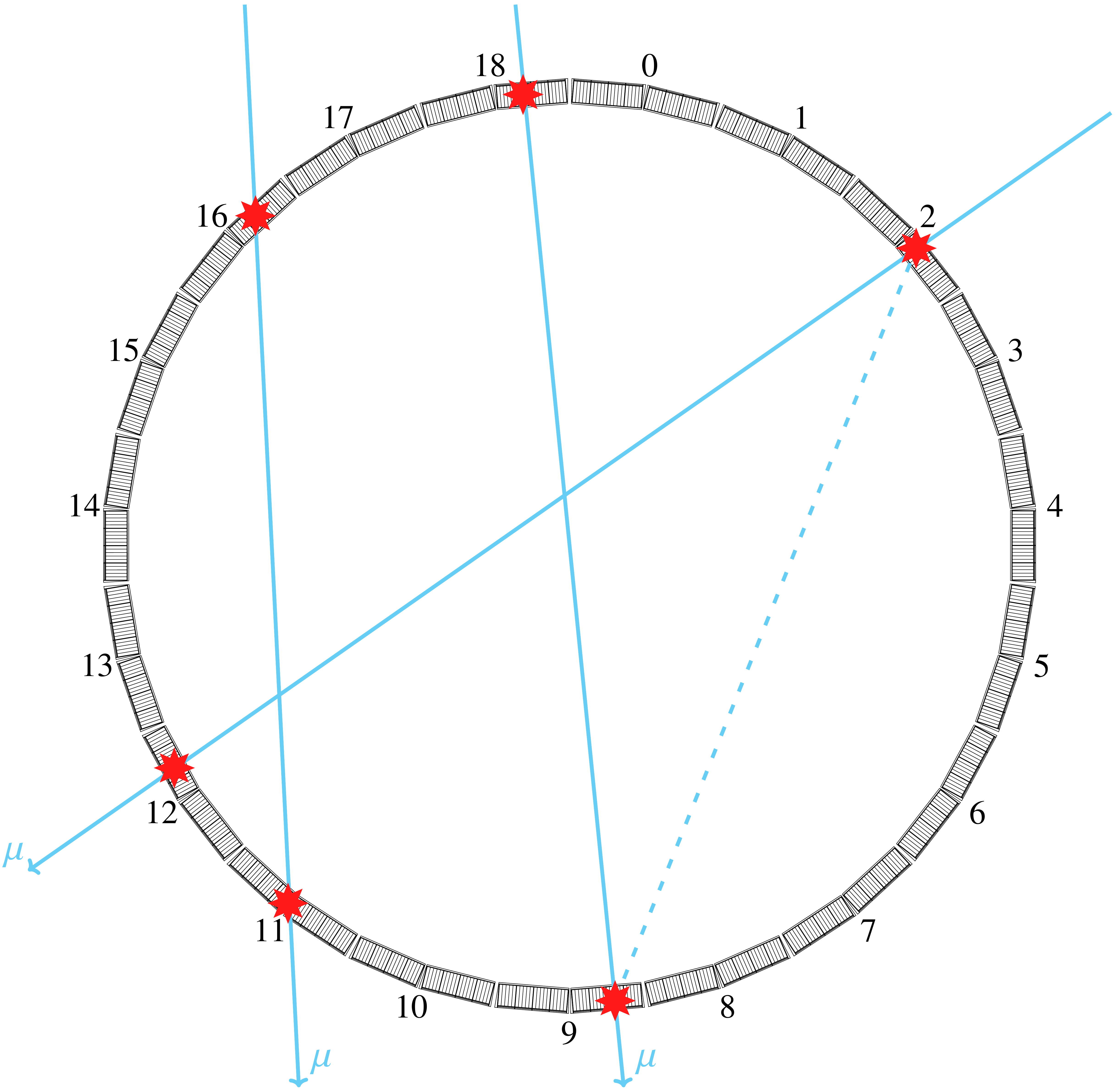}
    \caption{Transaxial sketch of Quadra with the LSO crystal arrays and numbered DEA. The red stars indicate single high-energy crystal hits of the muons crossing the gantry. The dashed line indicates a possible random coincidence.} \label{fig:gantrycrossing}
\end{figure}

Quadra is optimized for restricting the spatial location of the $e^+ e^- \rightarrow \gamma \gamma$ annihilation along the line of response of the two coincident photons through TOF \cite{Conti2019}. In order to exploit the TOF capability of Quadra, we select possible muon events by identifying coincident interactions across the gantry that meet the expected TOF for the muon. In addition, the coincident events must have energy values in the last energy bin. Fig.~\ref{fig:gantrycrossing} shows a transaxial view of Quadra with three muon crossing events. Selecting such crossing events requires us to identify the pairs of high-energy single crystal hits marked as red stars in Fig.~\ref{fig:gantrycrossing} in the singles mode data. 

First, we filter all high-energy crystal events, i.e.\ only the highest energy bin is considered. Then, we select the crossing events as follows: 

\begin{enumerate}
    \item Take the next single crystal event $h_0$ from the time-ordered list of high-energy single crystal events. 
    \item Create all possible pairs of candidate crossing events $(h_0,h_i)$ such that the second single crystal event $h_i$ fall within a wide time window relative to $h_0$, i.e.\ $\Delta t_1 \leq t_{h_i} - t_{h_0} \leq \Delta t_2$. 
    \item Qualify every candidate crossing event $(h_0,h_i)$ according to:  
    \begin{itemize}
        \item Restrict the Euclidean distance between $h_0$ and $h_i$: $|\vec{x}_{h_i} - \vec{x}_{h_0}| \leq d_{max} $. 
        \item Impose a minimal transaxial distance between $h_0$ and $h_i$: $|\vec{x}_{h_i} - \vec{x}_{h_0}|_{\perp\vec{y}} \geq d_{tr} $. 
        \item The crossing is in downward direction: $\cos\theta = \frac{(\vec{x}_{h_i} - \vec{x}_{h_0})\cdot \vec{z}}{|\vec{x}_{h_i} - \vec{x}_{h_0}|}  \leq 0$. 
        \item Require speed of light consistency, i.e.\ the distance and time difference between $h_0$ and $h_1$ must be consistent with an ultrarelativistic muon: $|1 - \frac{|\vec{x}_{h_i} - \vec{x}_{h_0}| }{ c \cdot (t_{h_i} - t_{h_0}) } |\leq r_c  $
    \end{itemize}
    \item Start again with the next single crystal event. 
\end{enumerate}
Tab.~\ref{tab:cuts} summarizes the parameter values of the cuts for the crossing event selection. This event selection was done off-line, i.e.\ not directly on the Quadra, with a dedicated prototype software written in Julia \cite{julia}.   

\begin{table}
    \centering
    \begin{tabular}{lc}
        \toprule
        Criterion & Value \\
        \midrule
        \rowcolor{gray!50}Single crystal event energy & $ E \geq \SI{726}{\keV}$ \\
        \multirow{2}*{Time window} & $\Delta t_1 = \SI{2.3}{\nano\second}$ \\
         & $\Delta t_2 = \SI{3.1}{\nano\second}$ \\
        \rowcolor{gray!50} Crossing direction & $\cos\theta \leq 0$ \\
        Maximal distance & $d_{max} = \SI{88}{cm}$ \\
        \rowcolor{gray!50} Minimal transaxial distance & $d_{tr} = \SI{80}{\cm}$  \\
        Speed of light consistency & $r_c = 0.005$ \\
        \bottomrule
    \end{tabular}
    \caption{Cuts on crossing event selection. }
    \label{tab:cuts}
\end{table}

As in normal PET coincidences (and also in the case of three-photon coincidences \cite{Steinberger2024}), our event selection algorithm inevitably leads to random crossing events, i.e.\ falsely paired single crystal events that do not belong to the same muon crossing the gantry (see dashed line in Fig.~\ref{fig:gantrycrossing}). Qualified random crossing events $(h_i,h_j)$ will have the wrong vector $\vec{x}_{h_i}-\vec{x}_{h_j}$ and therefore a wrong zenith angle $\theta$. Since we have no energy information beyond \SI{726}{\keV}, the only way to identify crossing events as muons truly crossing the scanner is through the angular distribution. As discussed in e.g.\ Ref.~\cite{PDG2020}, the angular distribution of muons at sea level follows a characteristic $\cos^2\theta$ distribution. We, therefore, consider the histogram of all measured $\cos \theta$ values for all qualified crossing events. The histograms of the $\cos \theta $ values will be fitted with a power law function 

\begin{equation}\label{eq:fitfunction}
    P(\cos\theta) \eq a_0 + a_1 \cdot \cos^q\theta \;, 
\end{equation}
with the three fitting parameters $a_0$, $a_1$ and $q$. 

The geometry of the gantry and the bias introduced by the crossing event selection algorithm on the angular distribution cannot be ignored \cite{Sullivan1971}. We simulated Quadra as a horizontally oriented cylinder with FLUKA \cite{fluka1,fluka2,fluka3,flair} and wrote a dedicated source routine that generates muons with a perfect $\cos^2 \theta$ distribution. A collision tape (MGDRAW routine) writes the spatial location of muons entering or exiting the cylinder. This allows for two checks: on the one hand we know to which history the muon belongs, i.e.\ each muon crossing the cylinder is tagged, and therefore we know exactly how many true muon crossing events we have in the simulation. This gives us control over the geometrical effects of the open cylindrical gantry of Quadra. On the other hand, we can simply run the crossing event selection algorithm on all registered single intersections of muon tracks and the cylinder (ignoring the particle history tag). This allows us to manage the crossing event selection bias and dependence on the parameter values in Tab.~\ref{tab:cuts}.

\section{Results}

\begin{figure}
    \centering
    \includegraphics[width=0.7\linewidth]{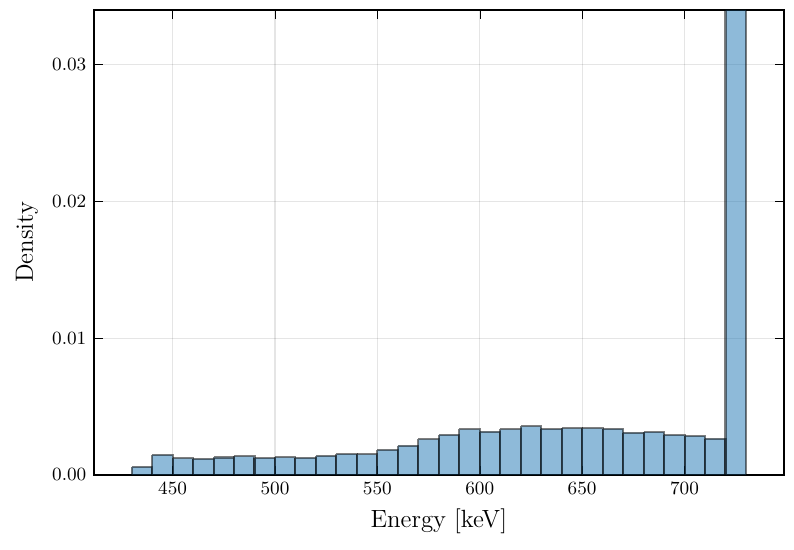}
    \caption{Energy spectrum of single crystal events recorded by Quadra in singles mode. }
    \label{fig:energyspectrum}
\end{figure}

First, Fig.~\ref{fig:energyspectrum} shows the energy spectrum of the blank scan without radioactive sources in the PET/CT scanner. The integrating energy bin is clearly visible. As a comparison, see Fig.~4 in Ref.~\cite{Mercolli2025} for the same energy spectrum but with a \isotope[44]{Sc} source inside the Quadra gantry. 

\begin{figure}
    \begin{subfigure}{0.52\textwidth}
        \includegraphics[width=\textwidth]{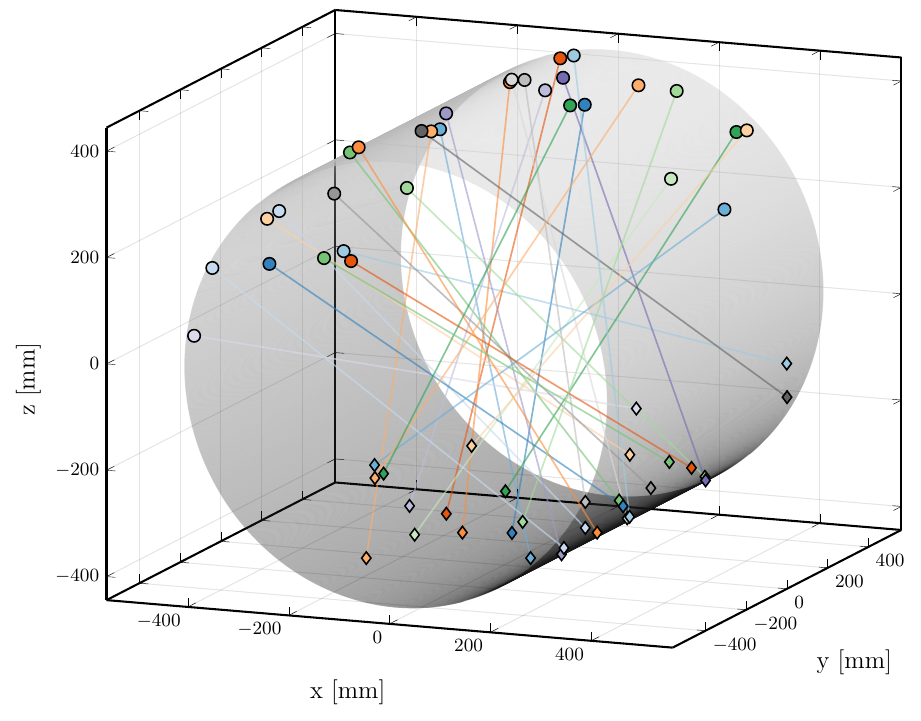}
        \caption{} \label{fig:crossingcandidates_3d}
    \end{subfigure}
    \hfill
    \begin{subfigure}{0.45\textwidth}
        \includegraphics[width=\textwidth]{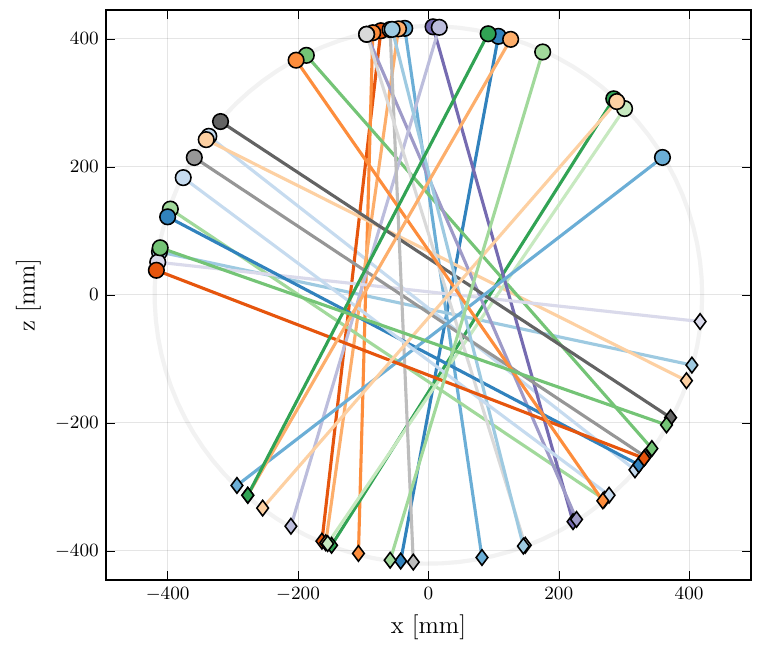}
        \caption{} \label{fig:crossingcandidates_2d}
    \end{subfigure}
    \caption{3D illustration (\subref{fig:crossingcandidates_3d}) and transaxial projection (\subref{fig:crossingcandidates_2d}) of 35 qualified crossing events in Quadra. The gray cylinder indicates the Quadra gantry. }
    \label{fig:crossingcandidates}
\end{figure}

Fig.~\ref{fig:crossingcandidates} illustrates 35 qualified crossing events from the measurement within the Quadra scanner. Note that these are not necessarily real muons crossing Quadra. We have no way to discriminate a true from a random crossing event on a single event basis.   

\begin{figure}
    \centering
    \includegraphics[width=0.7\linewidth]{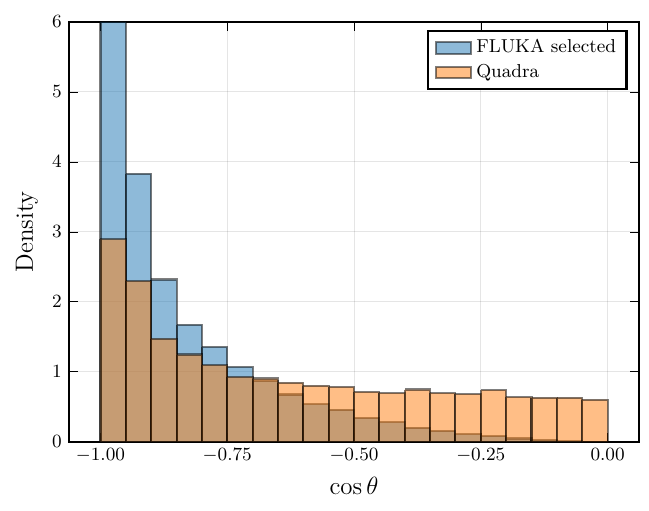}
    \caption{Angular distribution of the qualified crossing events from the Quadra data and the FLUKA simulation.}
    \label{fig:angulardistribution}
\end{figure}

The angular distributions of the qualified crossing events from the Quadra data and from the FLUKA data are shown in Fig.~\ref{fig:angulardistribution}. The histogram from the FLUKA data shows hardly any random events and the fitting parameter $a_0$ from Eq.~\eqref{eq:fitfunction} remains quite small $\hat{a}_0 =0.232 \pm 0.062 $. On the other hand, the histogram from the Quadra measurement clearly shows a constant amount of random crossing events that starts dominating around $\cos\theta \gtrsim -0.6$. The resulting point estimate for $a_0$ is $\hat{a}_0 =0.698 \pm 0.022 $.

Most importantly, fitting the power law of Eq.~\eqref{eq:fitfunction} to the distributions in Fig.~\ref{fig:angulardistribution} yields a point estimate for $q$ of $\hat{q}= 7.83 \pm 0.55$ for the FLUKA data and $\hat{q}=8.10 \pm 0.53$ for the measured data. Since the muons in the FLUKA simulation follow a perfect $\cos^2 \theta$ distribution, this means that apart from the random coincidences, the angular distribution of crossing events measured with Quadra is consistent with a $\cos^2\theta$ distribution.

\section{Discussion}

\begin{figure}
    \centering
    \begin{subfigure}{0.45\textwidth}
        \includegraphics[width=\linewidth]{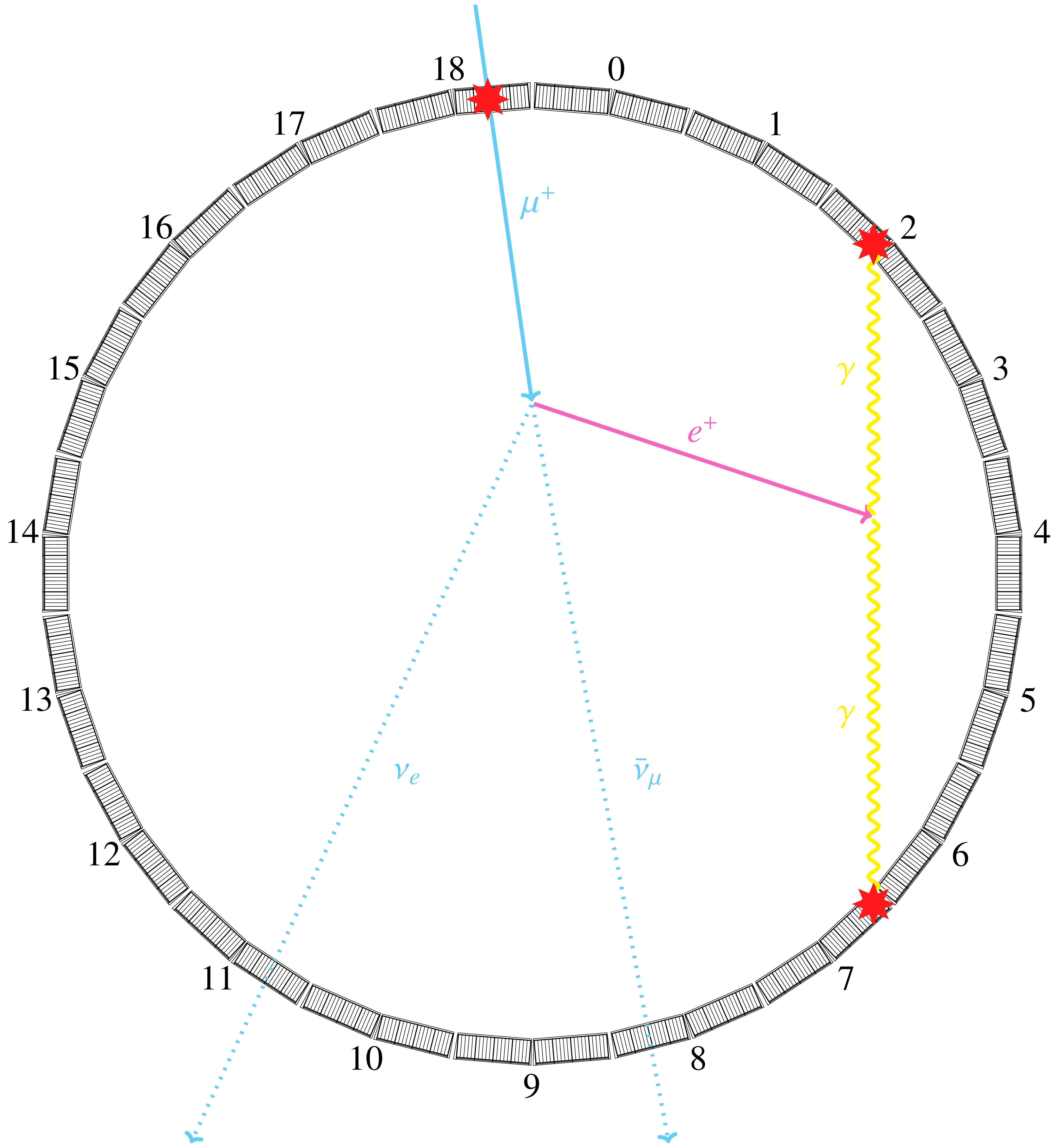}
        \caption{}\label{fig:muondecaysketch}
    \end{subfigure}
    \hspace{3cm} 
    \begin{subfigure}{0.095\textwidth}
        \includegraphics[width=\linewidth]{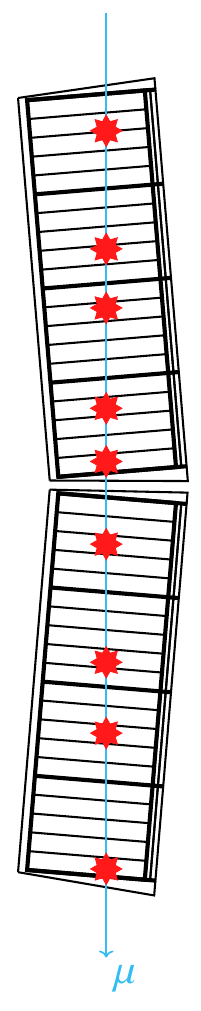}
        \caption{}\label{fig:singledeasketch}
    \end{subfigure}
    \caption{\subref{fig:muondecaysketch}: positive muon entering the Quadra detector ring and decaying into a a positron and a neutrino–antineutrino pair. \subref{fig:singledeasketch}: muon crossing a single DEA.}
    \label{fig:alternativeevents}
\end{figure}

The architectural detection constraints imposed by Quadra significantly impact the possible ways to detect atmospheric muons and play a key role in the event selection algorithm. An additional possible way to identify muon interactions would be to look for muons that enter Quadra and then decay before exiting the gantry. Fig.~\ref{fig:muondecaysketch} illustrates this process. This would generate the same signal as was used for positronium lifetime measurements with Quadra \cite{Steinberger2024,Mercolli2025a}, i.e.\ one high-energy single crystal event followed by two annihilation signals. This method did not lead to any useful result. Muon decays are quite rare given the size of the Quadra gantry and the amount of random three-event coincidences completely dominates the signal. 
Another possibility would be to look for muon tracks. A muon directly crossing a horizontal DEA (e.g.\ DEA no.\ 4) should leave a track of high-energy single crystal events as shown in Fig.~\ref{fig:singledeasketch}. However, we were unable to find multiple high-energy single crystal events within $\SI{0.45}{\ns}$ (about the time it takes a relativistic muon to cross a DEA), that would signal a muon track. This is likely due to the details of the localization of high-energy single crystal events, the scatter recovery and/or the event veto of Quadra. 

In order to fully exploit the properties of Quadra, in particular its TOF capabilities, we decided to select possible crossing gantry events by searching for two coincident high-energy single crystal events. This methodology, however, entails the presence of random coincidences (dashed line in Fig.~\ref{fig:gantrycrossing}). It is therefore also not possible to identify whether a single crossing event was generated by a muon or whether it is a random coincidence. We can only make a statement about an ensemble of qualified crossing events. In the absence of resolved crystal event energies above \SI{726}{\keV}, the only way to positively identify gantry crossing events as originating from atmospheric muons is therefore their characteristic $\cos^2\theta$ distribution.

The event selection algorithm is greedy, i.e.\ all qualified pairs $(h_0,h_i)$ are considered a crossing event. Inevitably, there will be random pairs (see the dashed line in Fig.~\ref{fig:gantrycrossing}). In addition to random crossing events from the selection algorithm, the natural abundance of \isotope[176]{Lu} in the LSO crystals can cause a fake single crystal event \cite{Steinberger2024}. In order to reduce the number of random crossing events, we have to impose the rather strict cuts reported in Tab.~\ref{tab:cuts}. The time window $\Delta t_1 ... \Delta t_2$ is the less relevant cut, i.e.\ it serves merely to restrict somewhat the number of $h_i$ events that can be paired with $h_0$, effectively speeding up the event selection algorithm. The distance limitations and speed of light consistency will essentially restrict the time difference between $h_0$ and $h_i$. The downward facing direction of travel $\cos\theta \leq 0$ as well as the consistency with the speed of light already greatly reduce the number of candidate crossing events. Note that $r_c = 0.005$ corresponds to a muon with $\approx \SI{1}{GeV}/c$, which is at the lower end of the muon spectrum at sea level (see Fig.~\ref{fig:muspectrum}). Considering that the diameter of Quadra is \SI{82}{\cm} and the maximal distance in the gantry is \SI{134.01}{\cm}, the values for $d_{tr}$ and $d_{max}$ in Tab.~\ref{tab:cuts} are chosen very restrictively. As a comparison: the minimal crystal distance for three-photon events in Ref.~\cite{Steinberger2024}, that effectively minimized the random coincidences from \isotope[176]{Lu}, is approximately \SI{100}{\mm}. However, loosening these two distance constraints leads to a large increase of the random crossing events. The histogram of Fig.~\ref{fig:angulardistribution} starts to peak at $\cos\theta > -1$ with a peak that depends on the cut value. 

The event selection algorithm with the cuts of Tab.~\ref{tab:cuts} leads to a significant steepening of the angular distribution of the Quadra crossing events: $\hat{q}=8.10 \pm 0.53$ when we naively would expect $q=2$ from the angular distribution of muons at sea level \cite{PDG2020}. Through a FLUKA simulation of muons with a $\cos^2\theta$ angular distribution, we confirm that the event selection and the cylindrical geometry of Quadra lead to this steepening of the angular distribution. Within the statistical uncertainty of the fit, the angular distribution from the Quadra data is consistent with the FLUKA data. Note that towards $\cos\theta \sim 0$ the FLUKA data shows few random crossing events, while the Quadra data flattens to a constant background value. We interpret these almost horizontal crossing events as stemming from the LSO, from high-energy secondaries (see Figs.~\ref{fig:deatracks_gamma} and \ref{fig:deatracks_e}) or naturally occurring gamma lines above \SI{726}{\keV} (e.g. from \isotope[40]{K} in the construction materials).  



\section{Conclusion}

We showed the first detection of atmospheric muons with a commercial PET/CT scanner. The limitations imposed by the properties of a PET/CT scanner could be circumvented, mostly by exploiting the TOF capabilities. A limitation of this approach is the dependence on angular distribution characteristics for event classification.

\newpage

\section*{Acknowledgements}

We thank Saverio Braccini, Maurizio Conti and Hasan Sari for useful discussions on the subject.

\section*{Declarations}

\subsection*{Funding}
\noindent This research is partially supported by the grant no. 216944 of the Swiss National Science Foundation.

\subsection*{Competing interests}
\noindent WMS is a full-time employee of Siemens Medical Solutions USA, Inc. LM has no conflict of interests to report.

\subsection*{Ethics approval}
\noindent Not applicable. 

\subsection*{Consent to participate}
\noindent Not applicable. 

\subsection*{Data availability}
\noindent The datasets generated and/or analyzed in the current study are available from the corresponding author upon reasonable request.

\subsection*{Authors' contributions}
\textbf{LM}: Conceptualization, Formal analysis, Investigation, Methodology, Software, Validation, Writing - Original Draft, Writing - Review \& Editing \newline 
\textbf{WMS:} Conceptualization, Investigation, Software, Writing - Review \& Editing


\newpage 
\bibliographystyle{elsarticle-num-names}
\biboptions{numbers,sort&compress}
\bibliography{muon_bibliography}

\end{document}